\documentclass[conference]{IEEEtran}
\IEEEoverridecommandlockouts

\usepackage{cite}
\usepackage{amsmath,amssymb,amsfonts}
\usepackage{algorithmic}
\usepackage{graphicx}
\usepackage{textcomp}
\usepackage{xcolor}

\usepackage{xurl}
\usepackage{hyperref}

\def\BibTeX{{\rm B\kern-.05em{\sc i\kern-.025em b}\kern-.08em
    T\kern-.1667em\lower.7ex\hbox{E}\kern-.125emX}}
\begin{document}

\begin{titlepage}
	\begin{center}	
		
		\huge
		\textbf{Dynamic Accuracy Estimation\\in a Wi-Fi-based Positioning System}
		
		\vspace{0.5cm}
		\LARGE
		Accepted version
		
		\vspace{1.5cm}
		
		\text{Marcin Kolakowski, Vitomir Djaja-Josko }
		
		\vspace{.5cm}
		\Large
		Institute of Radioelectronics and Multimedia Technology,\\Warsaw University of Technology,
		Warsaw, Poland,\\
		contact: marcin.kolakowski@pw.edu.pl

		\vspace{1.4cm}

	\end{center}
	
	\Large
	\noindent
	\textbf{Originally presented at:}
	
	\noindent
	2025 33rd Telecommunications Forum (TELFOR), Belgrade, Serbia
	
	\vspace{.5cm}
	\noindent
	\textbf{Please cite this manuscript as:}

	\noindent
	M. Kolakowski and V. Djaja-Josko, "Dynamic Accuracy Estimation in a Wi-Fi-Based Positioning System," 2025 33rd Telecommunications Forum (TELFOR), Belgrade, Serbia, 2025, pp. 1-4, doi: 10.1109/TELFOR67910.2025.11314159.
	
	\vspace{.5cm}
	\noindent
	\textbf{Full version available at:}
	
	\noindent
	\url{https://doi.org/10.1109/TELFOR67910.2025.11314159}

	\vspace{.5cm}
	\noindent
	\textbf{Additional information:}

	\noindent
	Associated data available at:
	
	\vspace{0.2cm}
	\noindent
	M. Kolakowski, ”Wi-Fi Fingerprints for Dynamic Accuracy Estimation Collected with a Robotic Platform,” Zenodo, Oct. 29, 2025. doi: \url{https://doi.org/10.5281/zenodo.17478483}.

	\vfill
	
	\large
	\noindent
	© 2025 IEEE.  Personal use of this material is permitted.  Permission from IEEE must be obtained for all other uses, in any current or future media, including reprinting/republishing this material for advertising or promotional purposes, creating new collective works, for resale or redistribution to servers or lists, or reuse of any copyrighted component of this work in other works.
\end{titlepage}

\title{Dynamic Accuracy Estimation\\in a Wi-Fi-based Positioning System

\thanks{This research was funded by the Polish National Centre for Research and Development, grant number THCS/I/24/RENEW/2025.}

}

\author{\IEEEauthorblockN{Marcin Kolakowski, Vitomir Djaja-Josko}
\IEEEauthorblockA{\textit{Institute of Radioelectronics and Multimedia Technology} \\
\textit{Warsaw University of Technology}\\
Warsaw, Poland \\
marcin.kolakowski@pw.edu.pl, vitomir.djaja-josko@pw.edu.pl}
}

\maketitle

\begin{abstract}
The paper presents a concept of a dynamic accuracy estimation method, in which the localization errors are derived based on the measurement results used by the positioning algorithm. The concept was verified experimentally in a Wi\nobreakdash-Fi based indoor positioning system, where several regression methods were tested (linear regression, random forest, k-nearest neighbors, and neural networks). The highest positioning error estimation accuracy was achieved for random forest regression, with a mean absolute error of 0.72 m.
\end{abstract}


\section{Introduction}

Positioning accuracy is one of the most critical parameters of   Real-Time Locating Systems (RTLS) as it directly influences the scope of services they can provide. For most systems, the information on accuracy is described in their datasheets with an averaged positioning error measured under certain conditions. Such information gives an overall impression of the system's performance, but does not allow for assessing the quality of single results.

Information on possible positioning errors can be helpful in many RTLS applications. For instance, in pedestrian localization systems, supplying the users with additional uncertainty information alongside the positioning result may allow them to interpret the result more easily. Such an approach is widely spread in smartphone applications using GNNS location data, where the uncertainty is marked with a circle around the computed location. In the case of autonomous systems, information on accuracy may be an additional variable in decision-making processes. Both applications require real-time accuracy estimation, which can be implemented with Dynamic Accuracy Estimation (DAE) methods.

The DAE methods aim to estimate the potential error of the computed object's location. Typically, the result is a radius of a circle, which contains the actual user location with assumed probability \cite{anagnostopoulosAnalysingDataDrivenApproach2021}. If the DAE methods differentiate the errors between the coordinate system axes, that area is elliptical \cite{valero-abundioUsingEllipsesImprove2024}. The positioning accuracy is usually estimated based on the same measurement results as used for the object's positioning \cite{lemicRegressionBasedEstimationLocalization2019}. Some methods consider additional parameters of the received signals, such as power level variation \cite{chenAnalysisPositioningAccuracy2020}.

There are two classes of DAE methods: rule-based and data-driven. In the rule-based algorithms, the accuracy is computed based on relationships between the measured signal parameters and the estimated location accuracy. The examples of such methods include algorithms based on Cramer-Rao Lower Bound analysis \cite{nikitinIndoorLocalizationAccuracy2017} and error propagation method \cite{valero-abundioUsingEllipsesImprove2024},  in which the measurement uncertainties are directly translated to the location accuracy.

The ongoing machine learning progress results in a rising interest in the data-driven methods. The method comparison presented in \cite{anagnostopoulosCanTrustThis2022} has proven that using the data-driven methods results in a higher quality of accuracy estimation than in the case of the rule-based ones. The data-driven DAE usually employs regression algorithms such as random forests or support vector machines  \cite{anagnostopoulosAnalysingDataDrivenApproach2021, lemicRegressionBasedEstimationLocalization2019}. For larger datasets, deep neural networks are also considered \cite{lemicArtificialNeuralNetworkbased2020}.

The paper presents an analysis of a data-driven DAE method's performance in a Wi--Fi-based positioning system, which was calibrated using an autonomous robotic platform. In the experiments, several machine learning methods were tested (k-nearest neighbors, random forest, neural networks, and linear regression) for error regression. The selected models were additionally tested on data gathered in a typical user positioning scenario.

\section{Dynamic Accuracy Estimation}

\subsection{Method's concept}
\label{sec:concept}

\begin{figure}[htbp]
	\centerline{\includegraphics[width=\linewidth]{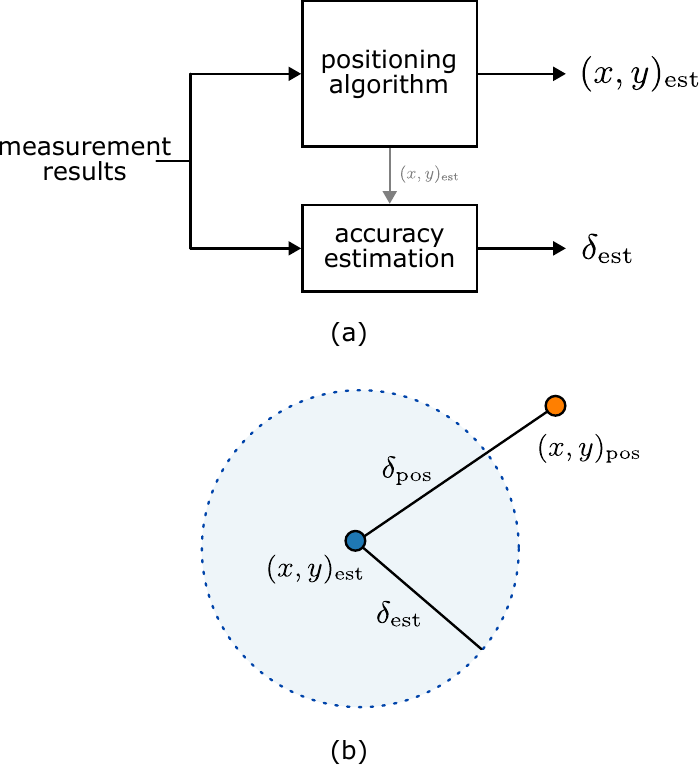}}
	\caption{DAE method concept; a) the algorithm workflow; b) relationship between the positioning result, true user location, and actual and estimated errors}
	\label{fig:general}
\end{figure}

The concept of DAE assumes implementing an accuracy estimation method that runs concurrently with the positioning algorithm. This idea is illustrated with Fig. \ref{fig:general}.a.

In the proposed concept, accuracy estimation is done based on the same measurement results as those used for positioning. An additional variant, where the positioning result is used, was also tested in the experiments. The result of the DAE is the estimated error value $\delta_{\rm{est}}$. The relationship between the positioning result $(x,y)_{\rm{est}}$, and the actual user location $(x,y)_{\rm{pos}}$ is presented in Fig. \ref{fig:general}. b.

The obtained error estimate $\delta_{\rm{est}}$ can be either smaller or larger than the actual error $\delta_{\rm{pos}}$. In the case of systems where the accuracy estimation is a basis for crucial decisions, the DAE algorithm parameters should be adjusted to take that into account.

\subsection{Wi-Fi based system implementation}
\label{sec:wifi_impl}

The described DAE concept was implemented in a Wi\nobreakdash-Fi, smartphone-based positioning system. In the system, the user is located based on power measurements of signals transmitted by  Wi\nobreakdash-Fi access points operating in 2.4 and 5 GHz bands. The user's location is derived through k-Nearest Neighbors (kNN) based fingerprinting (k=4). Error estimation is performed using regression methods trained on a dedicated dataset.

\begin{figure}[htbp]
	\centerline{\includegraphics[width=.55\linewidth]{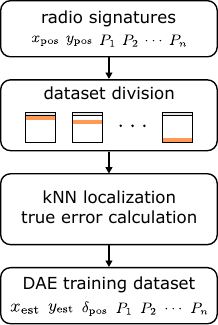}}
	\caption{The process of preparing the DAE training dataset.}
	\label{fig:dataset}
\end{figure}

The dataset for DAE methods training should reflect the relationship between the parameters of the received signals and true positioning errors in the system and its environment. The procedure for the dataset preparation is presented in Fig.~\ref{fig:dataset}.

The initial dataset includes radio signatures (measured power levels of signals received from the access points, annotated with the measurement location). The initial dataset is partitioned several times into training and testing sets according to the k-fold cross-validation principles (every time, the test set includes unique radio signatures). For each partition, the training set is treated as a fingerprinting radio map which is used to compute locations of the measurement points from the test subsets using kNN. For each location estimate ($x_{\rm{pos}}, y_{\rm{pos}}$), a true positioning error $\delta_{\rm{pos}}$ is calculated. The obtained values are combined with the power level measurements, creating a piece of the DAE training dataset. Adopting the described approach enables the use of all of the collected signatures for DAE training, which is especially important in the case of sparse maps, where the number of the signatures is limited.

\section{Experiments}
\subsection{Experimental environment}

The described concept was experimentally tested in office and laboratory rooms of the Faculty of Electronics and Information Technology of Warsaw University of Technology. The general layout is presented in Fig. \ref{fig:poses}. The radio signal propagation conditions in the experiment location are harsh due to a large amount of equipment collected in the rooms and metal elements in partition walls. 

The experiments consisted in acquiring the dataset using an autonomous robotic platform and then using it to train DAE data-driven methods.

\begin{figure}[htbp]
\centerline{\includegraphics[width=.8\linewidth]{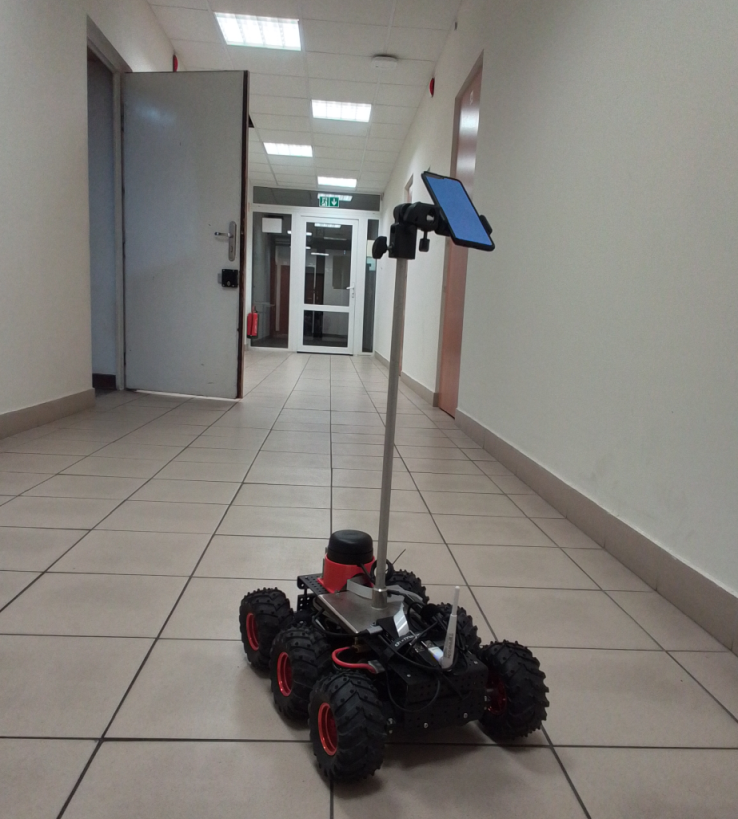}}
\caption{The robotic platform in the experiment environment}
\label{fig:robot}
\end{figure}

\begin{figure}[htbp]
\centerline{\includegraphics[width=.9\linewidth]{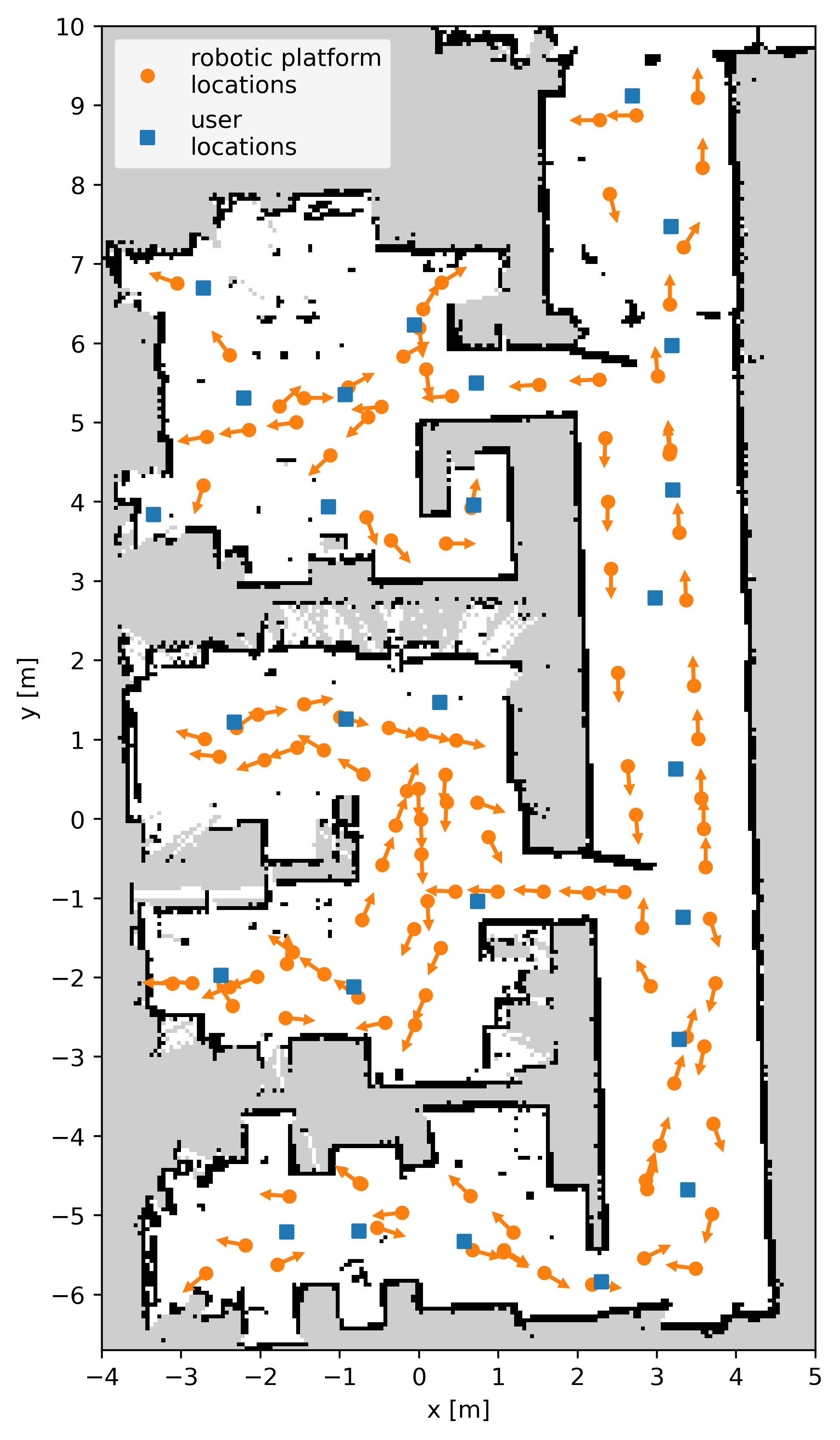}}
\caption{Measurement points' locations. The arrows represent the platform's orientation.}
\label{fig:poses}
\end{figure}

During the experiments, a smartphone (Samsung A41) was mounted on a stand that was fixed to a robotic platform. A photo of the setup is shown in Fig. 3.  The smartphone's elevation was 0.74 m above the ground.
The reference platform location was obtained with a cm-level accuracy using a LiDAR sensor (RPLidar A2M12) with a nominal range of 12 m. The whole platform was controlled using the libraries and tools of the Robot Operating System (ROS) suite.

An additional experiment, in which the user held the smartphone, was performed to verify the method's performance under real operational conditions.

\subsection{Data sets}

The primary dataset comprised 359 radio signatures (Wi-Fi signal levels, reference smartphone location) collected in 117 points (minimum three scans per point) using the robotic platform. The locations of the measurement points are presented in Fig. \ref{fig:poses}. In the experimental area, 78 different access points were detected. As some of them were distant and their signals were received in only a few points, the dataset was built using data from the 35 most available anchors. To make the dataset usable by all algorithms, the missing power measurements were replaced with a constant value of -99 dBm (lower than the smartphone's sensitivity). The DAE training dataset was prepared according to the procedure described in Section \ref{sec:wifi_impl}. For the given experimental setup, the average robot positioning error equaled 0.97 m.

The second dataset comprised 108 signatures collected by the user in 27 points marked in Fig. \ref{fig:poses}. The average user positioning error was relatively high (2.49 m) because he was localized based on robot-obtained fingerprinting map. This dataset was used only to evaluate the performance of DAE models trained on robot-obtained data.
Both datasets are publicly available online on Zenodo \cite{Github}.

\subsection{DAE performance evaluation}

To evaluate the described concept, several machine learning methods were used: Linear Regression (LR), Random Forest (RF), k-Nearest Neighbors (kNN) and Dense Neural Networks (NN). They were trained on the  DAE dataset prepared as described in Section \ref{sec:wifi_impl}. All methods, besides NNs, were implemented using the scikit-learn suite. Neural networks were implemented with PyTorch. Prior to the final evaluation, several architectures consisting of dense layers followed by batch normalization were tested. The presented results are the best that were obtained (the number of neurons in specific layers is listed as a parameter in Table \ref{tab1}).

\begin{figure}[htbp]
\centerline{\includegraphics[width=.95\linewidth]{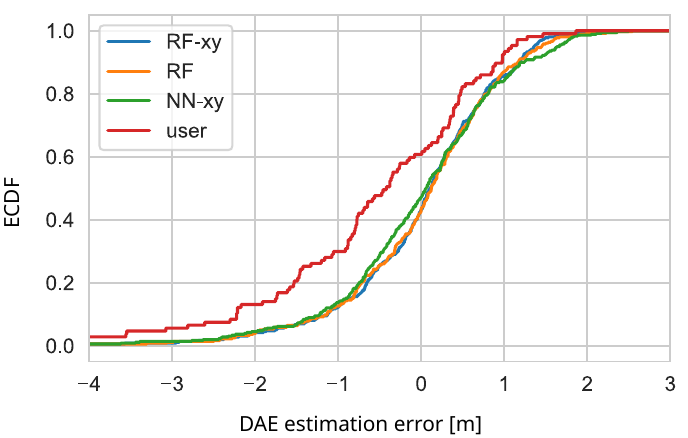}}
\caption{Empirical cumulative distribution of the signed DAE error}
\label{fig:signed_cdf}
\end{figure}

\begin{figure}[htbp]
	\centerline{\includegraphics[width=.95\linewidth]{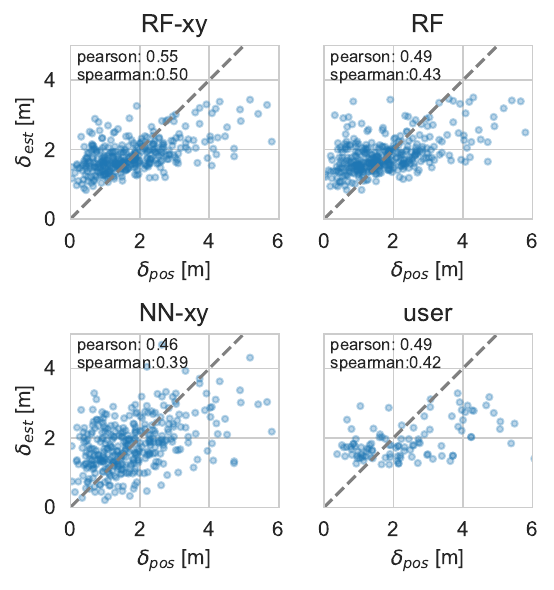}}
	\caption{Correlation between true and estimated positioning errors}
	\label{fig:corrs}
\end{figure}

The approach, in which the computed user location is taken into account, was also tested (the results for this scheme are marked with a -xy suffix). The performance of methods was compared based on the DAE error defined as:
\begin{equation}
\rm{DAE_{error}} = \delta_{\rm{est}} - \delta_{\rm{pos}}
\end{equation}
where $\delta_{\rm{est}}$ and $\delta_{\rm{pos}}$ are defined by Fig. \ref{fig:general}. The values of the mean absolute error (MAE) and the mean squared error (MSE) with the method's parameters are collected in Table~\ref{tab1}. The empirical cumulative distribution function of the DAE error and the relationship between the real and estimated error values are presented in Fig. \ref{fig:signed_cdf} and Fig. \ref{fig:corrs}, respectively.

In all cases, considering the estimated smartphone location leads to increased DAE quality. The highest accuracy was obtained for the random forest (mean MAE of 0.73 m). This method was tested on the second, user-collected dataset. For user-collected data, the accuracy estimation errors were higher due to different propagation conditions in which the data were gathered (the robot-collected data do not account for body shadowing). 

\begin{table}[htbp]
\caption{Mean errors of dynamic accuracy estimation\\of robot's and users location}
\begin{center}
\begin{tabular}{|c|c|c|c|}
\hline

		\textbf{algorithm} & \textbf{parameters} & \textbf{MAE}& \textbf{MSE}\\ 
		\hline
		LR		& - 		& 0.926 & 1.529 \\
		LR-xy &	 - 			& 0.921 & 1.511 \\
		RF &	trees=100	& 0.753 & 1.046 \\
		RF-xy &	trees=300	& 0.726 & 0.980 \\
		kNN & k=4 			& 0.831 & 1.295 \\
		kNN-xy & k=4 		& 0.813 & 1.261 \\
		NN & [128, 128, 128]			& 0.859 & 1.283 \\
		NN-xy &  [256, 512, 256]			& 0.806 & 1.135 \\
		\hline
		user & RF-xy (300) &1.082 & 2.112\\
		\hline

\end{tabular}
\label{tab1}
\end{center}
\end{table}

In the results obtained for the robot positioning error estimation, we did not observe any tendency to either overestimate or underestimate the estimated errors. However, such a tendency exists in the case of user positioning, where the estimated errors are more optimistic than in reality. In case of the tested methods, the obtained estimates show moderate correlation with the actual values. The Pearson correlation coefficient is around 0.5, close to the result obtained with other methods~\cite{anagnostopoulosCanTrustThis2022}.

\section{Conclusions}

The paper presents a method for dynamic estimation of positioning accuracy in a Wi-Fi-based system. The technique consists in training a machine learning model which, based on the measurement results, assesses the accuracy of the computed locations. The experimental verification showed that the method can be successfully used for positioning error estimation. However, the quality of estimation deteriorates when the system is used in drastically different propagation conditions (e.g. when the device is held by the user rather than being on a stand). In such situations, model adaptation might be required.

The proposed conception shows potential for further development. One of the possible directions is considering other signal parameters or using loss functions to penalize estimates that are too optimistic.

\end{document}